\pdfoutput=1
\documentclass[prl,twocolumn,superscriptaddress,preprintnumbers,nofootinbib]{revtex4}
\usepackage{graphicx}
\usepackage{epsfig}
\usepackage{bm}
\usepackage{latexsym,amssymb,amsmath,amsfonts,amssymb,txfonts,wasysym,float}
\usepackage{scrextend}
\usepackage{mathrsfs}
\usepackage{color}

\usepackage{enumitem}

\usepackage[usenames,dvipsnames]{xcolor}
\definecolor{orange}{cmyk}{0,0.5,1,0}
\definecolor{rossoCP3}{cmyk}{0,.88,.77,.40}
\definecolor{graa}{rgb}{0.8,0.8,0.8}
\definecolor{blaa}{rgb}{0.2,0.2,0.6}

\begin{document}
\preprint{MPP-2020-121}
\preprint{LMU-ASC 34/20}

\title{\color{rossoCP3} Anomalous $\bm{U(1)}$ Gauge Bosons as Light
  Dark Matter in String Theory}

\author{Luis A. Anchordoqui}

\affiliation{Department of Physics and Astronomy,  Lehman College, City University of
  New York, NY 10468, USA
}

\affiliation{Department of Physics,
 Graduate Center, City University
  of New York,  NY 10016, USA
}

\affiliation{Department of Astrophysics,
 American Museum of Natural History, NY
 10024, USA
}

\author{Ignatios Antoniadis}
\affiliation{Laboratoire de Physique Th\'eorique et Hautes \'Energies - LPTHE
Sorbonne Universit\'e, CNRS, 4 Place Jussieu, 75005 Paris, France
}

\affiliation{Albert Einstein Center, Institute for Theoretical Physics
University of Bern, Sidlerstrasse 5, CH-3012 Bern, Switzerland
}

\author{Karim Benakli}

\affiliation{Laboratoire de Physique Th\'eorique et Hautes \'Energies - LPTHE
Sorbonne Universit\'e, CNRS, 4 Place Jussieu, 75005 Paris, France
}

\author{Dieter\nolinebreak~L\"ust}

\affiliation{Max--Planck--Institut f\"ur Physik,  
 Werner--Heisenberg--Institut,
80805 M\"unchen, Germany
}

\affiliation{Arnold Sommerfeld Center for Theoretical Physics 
Ludwig-Maximilians-Universit\"at M\"unchen,
80333 M\"unchen, Germany
}

\begin{abstract}
  \vskip 2mm \noindent Present experiments are sensitive to very
  weakly coupled extra gauge symmetries which motivates further
  investigation of their appearance in string theory compactifications
  and subsequent properties. We consider extensions of the standard
  model based on open strings ending on D-branes, with gauge bosons
  due to strings attached to stacks of D-branes and chiral matter due
  to strings stretching between intersecting D-branes. Assuming that
  the fundamental string mass scale saturates the current LHC limit and that the theory is weakly coupled, we
  show that (anomalous) $U(1)$ gauge bosons which propagate into the
  bulk are compelling light dark matter candidates. We comment on the
  possible relevance of the  $U(1)$ gauge bosons, which are universal in  intersecting D-brane models, to the observed $3\sigma$
  excess in XENON1T.
\end{abstract}
\maketitle

\section{Introduction}
The primary objective of the High Energy Physics (HEP) program is to
find and understand what physics may lie beyond the Standard
$S\!U(3)_C\otimes S\!U(2)_L \otimes U(1)_Y$ Model (SM), as well as its
connections to gravity and to the hidden sector of particle dark
matter (DM). This objective is pursued in several distinct ways. In
this Letter, we explore one possible pathway to join the vertices of
the HEP triangle using string compactifications with large extra
dimensions~\cite{Antoniadis:1998ig}, where sets of D-branes lead to
chiral gauge sectors close to the
SM~\cite{Blumenhagen:2005mu,Blumenhagen:2006ci}.

D-branes provide a nice and simple realization of non-abelian gauge
symmetry in string theory. A stack of $N$ identical parallel D-branes
eventually generates a $U(N)$ theory with the associated $U(N)$ gauge
group where the corresponding gauge bosons emerge as excitations of open
strings ending on the D-branes. Chiral matter is either due to strings
stretching between intersecting D-branes, or to appropriate
projections on strings in the same stack. Gravitational interactions are described by closed
strings that can propagate in all dimensions; these comprise parallel
dimensions extended along the D-branes and transverse ones.

String compactifications could leave characteristic footprints at particle colliders:
\begin{itemize}[noitemsep,topsep=0pt]
\item the emergence of Regge recurrences at parton collision energies
  $\sqrt{s} \sim$ string mass scale $\equiv M_s = 1/\sqrt{\alpha'}$~\cite{Anchordoqui:2007da,Anchordoqui:2008di,Anchordoqui:2014wha};
\item the presence of one or more additional $U(1)$ gauge symmetries, beyond the $U(1)_Y$ of the SM~\cite{Ghilencea:2002da,Anchordoqui:2011eg,Cvetic:2011iq}.
\end{itemize}
Herein we argue that the (anomalous) $U(1)$ gauge bosons that do not
partake in the hypercharge combination could become compelling dark matter candidates. Indeed, as noted
elsewhere~\cite{Antoniadis:2002cs} these gauge fields could
live in the bulk and the four-dimensional $U(1)$ gauge coupling would
become infinitesimally small in low string scale models, $g \sim M_s /M_{\rm Pl}$,
where $M_{\rm Pl}$ is the Planck mass (for previous investigations in different regions of parameters and different string scenarios, see for example \cite{Abel:2008ai,Burgess:2008ri,Goodsell:2009xc}). Note that for typical energies $E$ of
the order of the electron mass, the value of $g$ is still bigger that
the gravitational coupling $\sim E/M_{\rm Pl}$, and the strength of the
new force would be about $10^7$ times stronger than gravity, where we have taken
$M_s \sim 8~{\rm TeV}$, saturating the LHC bound~\cite{Sirunyan:2019vgj}.

To develop some sense for the orders of magnitude involved, we now
  make contact with the experiment. The XENON1T Collaboration has
  recently reported a surplus of events in
  $1 \alt {\rm electronic\ recoils}/{\rm keV} \alt
  7$, peaked around 2.8~keV~\cite{Aprile:2020tmw}. The total number of events recorded within
  this energy window is 285, whereas the expected background is
  $232 \pm 15$. Taken at face value this corresponds to a significance
  of roughly $3\sigma$, but unknown backgrounds from tritium decay
  cannot be reliably ruled out~\cite{Aprile:2020tmw}. Although the excess is not statistical
  significant, it is tempting to imagine that it corresponds to a real
  signal of new physics. A plethora of models have already been
  proposed to explain the excess, in which
  the DM particle could be either the main component of the abundance in the solar
neighborhood, $n_{\rm DM} \sim 10^5  \left(m_{\rm DM}/2.8~{\rm
    keV}\right)^{-1}~{\rm cm}^{-3}$,
or else a sub-component of the DM
population. Absorption
of a $\sim 2.8~{\rm keV}$ mass dark vector boson that saturates the local DM
mass density provides a good fit to the
excess for a $U(1)_X$ gauge coupling to electrons of $g_{X,{\rm eff}} \sim 2 \times 10^{-16} - 8 \times 10^{-16}$~\cite{Aprile:2020tmw,Benakli:2020vng,Alonso-Alvarez:2020cdv,Choi:2020udy,An:2020bxd,Okada:2020evk}. For such small masses and couplings, the
cosmological production should be non-thermal
\cite{Alonso-Alvarez:2020cdv}, avoiding  constraints from structure
formation~\cite{Gilman:2019nap,Banik:2019smi}. Leaving aside attempts
to fit the XENON1T excess, we might consider a wider range of dark
photon masses and couplings. For light and very weakly coupled dark
photons, the cooling of red giants and horizontal branch stars give stronger or similar bounds on $g_{X,{\rm
    eff}}$ than direct detection
experiments~\cite{An:2013yfc,Fabbrichesi:2020wbt}.\footnote{A point
  worth noting at this juncture, however, is that there are several stellar
systems that exhibit a mild preference for 
an over-efficient cooling mechanism when compared to theoretical
models~\cite{Giannotti:2015kwo}. Thus, the argument can be turned around and the anomalous cooling  could be interpreted as evidence for $U(1)_X$ production in  dense star cores.} For instance, rescaling the bounds
quoted in~\cite{An:2020bxd} leads to an upper bound $g_{X, {\rm eff}}
\gtrsim 10^{-16} - 10^{-14}$ for $m_X$ varying from $10$ to $100$
keV. As an example, if we take  $m_X \sim 15~{\rm keV}$ in agreement
with the bound of  $\sim 5$~keV ~\cite{Gilman:2019nap,Banik:2019smi},
the upper bound is about $g_{X,{\rm eff}} \lesssim 5 \times10^{-16}$. Obtaining such small values of masses and couplings for the dark photon are challenging as we will show.
 
 \section{Generating the small $U(1)_X$ couplings}
 
 \subsection{Open string models}
We start from ten-dimensional type I string theory compactified on a six-dimensional space of volume $V_6 M_s^6$. The relation between the
Planck mass, the string scale, the string coupling $g_s$,
and the total volume of the bulk $V_6 M_s^6$ reads:
\begin{equation}
 M_{\rm Pl}^2 = \frac{8}{g_s^2} \ M_s^8 \ \frac{V_6}{(2 \pi)^6} \, .
\label{mpl}
\end{equation}
A hierarchy between the Planck and string scales can be due to either a large volume $V_6 M_s^6 \gg 1$ or a very small string coupling. 
We discuss these two possibilities successively.

From now on, we denote by $d$ the total number of dimensions that are large.  For simplicity, we assume that they have a common radius $R$ while the other $6-d$ dimensions have a radius $M_s^{-1}$. Obviously, the latter have no more a classical geometry and supergravity description as space dimensions; they represent new degrees of freedom that have a stringy description, for example through the corresponding world-sheet conformal field theories. The couplings of all light states are under control and our formulae still hold.
The $U(1)_X$ gauge fields live on a  D$(3+\delta_X)$-brane that wraps a $\delta_X$-cycle of volume $V_X$, while its remaining
four dimensions extend into the uncompactified space-time. The corresponding gauge coupling is given by:
\begin{equation}
    g_X^2 =\frac{(2 \pi)^{\delta_X +1} \ g_s}{V_X \ M_s^{\delta_X}} \,.
\label{gb}
  \end{equation} 
Assuming all the $\delta_X$-cycles are  sub-spaces of  internal $d$ large dimensions with the same radius, the substitution of (\ref{mpl}) into (\ref{gb}) leads to:
\begin{equation}
g_X^2 = 2 \pi g_s \left( \frac{8}{g_s^{2}} \right)^{\delta_X/d}  \ \left(\frac{M_s}{M_{\rm
        Pl}} \right)^{2 \delta_X/d} \, .  
        \end{equation}

It is straightforward to see that to realize the weakest gauge interaction the volume seen by $U(1)_X$
must exhaust the total large internal volume suppressing the strength of gravitational interactions $\delta_X=d$ (as in \cite{Benakli:2020vng}), yielding
\begin{equation}
g_X = \sqrt{\frac{16 \ \pi}{g_s}} \ \frac{M_s}{M_{\rm Pl}}
\sim  4 \times10^{-14}   \, \, {\left( \frac{0.2}{g_s} \right)^{1/2}} \left(\frac{M_s}{10 \, \, {\rm  TeV}} \right)\,,
\label{coupling}
\end{equation}
where we have taken as reference values $g_s = 0.2$ and $M_s \gtrsim 10$ TeV. The latter is a conservative bound from non-observation of stringy excitations 
at colliders ~\cite{Sirunyan:2019vgj} while a slightly stronger bound of order, but model dependent, can be obtained from limits on dimension-six four-fermion operators~\cite{Accomando:1999sj,Cullen:2000ef,Antoniadis:2000jv,Giudice:2003tu}. As for $g_s$ we will consider that it is in the range $0.01 - 0.2$, and could be fixed after a careful study of running of the gauge couplings. In the case of toroidal compactifications,  the internal six-dimensional volume is expressed in terms of the parallel and transversal radii as
\begin{equation}
 V_6 = (2\pi)^6 \ \prod_{i=1}^{d_\parallel} \
  R_i^\parallel \
  \prod_{j=i}^{d_\perp} R_j^\perp \, ,
\end{equation}
where now for each stack of Dp-branes we identify the corresponding $d_\parallel = \delta$.  For instance, if the SM  arise from  D3-branes and the $U(1)_X$ from  D7-branes with an internal space having four large dimensions all parallel to the D7-brane world-volume ($\delta_X=d=4$), we get for $g_X$ the result in (\ref{coupling}).

 \subsection{Little strings models}

Another possibility for engineering extremely weak extra gauge symmetries is to consider a scenario which allows very small value of $g_s$. Such a possibility is provided by small instantons~\cite{Witten:1995gx,Benakli:1999yc} or Little String Theory (LST)~\cite{Antoniadis:2001sw,Antoniadis:2011qw} where we localize the SM gauge group on Neuveu-Schwarz (NS) branes (dual to the D-branes).

In the case of LST~\cite{Antoniadis:2001sw,Antoniadis:2011qw}, we start with a compactification on a six-dimensional space of volume $V_6$ with the Planck mass given by (\ref{mpl}) (up to a factor 2 in the absence of an orientifold). The internal space is taken as a product of a two-dimensional space, of volume $V_2$, times a four-dimensional compact space, of volume $V_4$.  However, instead of D-brane discussed above, we assume that the SM degrees of freedom emerge on a stack of NS5-branes wrapping the two-cycle of volume $V_2$. We take for simplicity this to be a torus made of two orthogonal circles with radii $R_1$ and $R_2$. The corresponding (tree-level) gauge coupling is given by:
\begin{equation}
g_{\rm SM}^2 = \frac{R_1}{R_2} \quad ({\rm Type \, \,  IIA}) \quad {\rm and}
\quad g_{\rm SM}^2 = \frac{1}{R_1 R_2 M_s^2}  \quad ({\rm Type \,  \,  IIB}); 
\label{gSM-LST}
\end{equation}
thus, an order one SM coupling imposes $R_1\simeq  R_2 \simeq  M_s^{-1 }$. On the other hand, the $U(1)_X$ is supposed to appear in the bulk and has a coupling
given by (\ref{gb}). If $U(1)_X$ arises from a D9-brane then: 
\begin{equation}
 M_{\rm Pl}^2  = \frac{8}{g_s^2} \ M_s^8 \ \frac{V_2 V_4}{(2 \pi)^6} \, ,\qquad {\rm and} \qquad   g_X^2 =\frac{(2 \pi)^{7} \ g_s}{V_2 V_4 \ M_s^6} \,.
\label{mpl-LST}
\end{equation}
Now, taking all the internal space radii to be of the order of the string length, $M_s^6 V_2 V_4 \simeq (2 \pi)^6$,  leads to:
\begin{equation}
g_{X}\simeq \ \sqrt{32} \pi \sqrt{\frac{M_s}{M_{\rm Pl}}} \sim 5\times
10^{-7} \left(\frac{M_s}{10 \, \, {\rm  TeV}} \right)^{1/2} \, .
\label{LST-coupling-X}
\end{equation}
Note however that the $U(1)_X$ from a D-brane does not interact directly with the electrons of the SM on the NS5-brane. Such interaction could arise via a closed string exchange which is likely to be suppressed by two powers of the string coupling, leading to an effective interaction of the order of $10^{-14}$.

 \subsection{Small instanatons models}

In heterotic strings compactified on $K3$, of volume
$V_{K3}$ fibered over a two-dimensional base $P^1$ of volume $V_{P^1}$
with integrated volume $<V_{K3} V_{P^1}>$, the Planck mass reads:
\begin{equation}
 M_{\rm Pl}^2 = \frac{64 \ \pi}{g_s^2} \ M_s^8 \ <V_{K3} V_{P^1}> \, . 
\label{mpl-H}
\end{equation}
Taking the limit of instanton small size leads to emergence of a gauge  group, identified with the SM one, supported at particular points on $K3$ . The corresponding gauge coupling reads:
\begin{equation}
g_{\rm SM}^2= \frac{ 2 \pi^2}{M_s^2 <V_{P^1}>} \,,
\label{Het-coupling-Instanton}
\end{equation}
implying that to give phenomenologically acceptable values, the compactification radius should remain of order of the string scale.  The $U(1)_X$ is identified within the bulk theory descending from the ten-dimensional gauge symmetry:
\begin{equation}
g_{X}= \frac{g_s} {2} \frac{ 1}{M_s^3 \sqrt{<V_{K3} V_{P^1}>}} =  4
\sqrt{\pi}  \frac{M_s}{ M_{\rm Pl}} \sim 6 \times 10^{-14} \frac{M_s}{
  10\ {\rm TeV}} \, .
\label{Het-coupling-X}
\end{equation}
Taking $<V_{K3} V_{P^1}> \simeq <V_{K3}>< V_{P^1}>$, we see that the weakness of gravitational interactions, and a consequence of the $U(1)_X$ coupling, can be due either to a large volume of the $K3$ internal space or to a small string coupling: $<V_{K3} >^{1/4} \sim $ GeV$^{-1}$ or $g_s \sim 10^{-13}$ for $M_s \sim 10$ TeV.

 \section{Dark photon mass generation}

We turn now to the generation of a mass for the dark photon. 

 \subsection{Higgs mechanism}

Let's denote by $v_X$ the vacuum expectation value for the Higgs $h_X$ that breaks the $U(1)_X$ symmetry. The simplest quartic potential $-\mu_X^2 h_X^2 + \lambda_X h_X^4$ leads to $v_X = {\mu_X}/\sqrt{{2 \lambda_X}}$, a Higgs mass of order $\mu_X$ and a mass for the  dark photon 
\begin{equation}
m_X =  \frac{g_X  \mu_X}{\sqrt{{2 \lambda_X}}} =   \sqrt{2 \pi g_s} \left(
  \frac{8}{g_s^2} \right)^{\delta/2 d} \  \left(\frac{M_s}{M_{\rm Pl}}
\right)^{\delta/d} \, \,  v_X \, .
\label{mass-Higgs-delta}
\end{equation}
This gives for $d=\delta=6$:
\begin{equation}
m_X \sim  \, \, {\left( \frac{0.2}{g_s} \right)^{1/6}} \left(\frac{M_s}{1000 \, \, {\rm TeV}} \right)^2  
\left(\frac{v_X} {M_s}\right)  {\rm keV} \, .
\label{mass-Higgs-6-dimensions}
\end{equation}
Taking $v_X \simeq M_s$, this leads to a mass of order $ 0.1$ to $1.4
\times10^3$~eV when varying $M_s$ from $10$ to $1000$~TeV, and $g_s$
from $0.2$ to $0.02$.  For this region of the parameter space, the gauge coupling
varies in the range $4 \times 10^{-14} \alt g_X \alt 2 \times 10^{-11}$. Higher photon masses are of course easier to obtain with smaller number of $d_\parallel$ dimensions. For example, an $M_s \sim 10$~TeV,  and  $M_s \sim 100$~TeV lead respectively to $m_X \sim 6$ keV, $g_X \sim  6 \ \times 10^{-10} $,  and $m_X \sim 270$ keV,  $g_X \sim  6 \ \times 10^{-9} $ for $\delta_X =4$, $d=6$ and $g_s \sim 0.2$  .

  \subsection{St\"uckelberg mechanism}

Another possibility is that the abelian gauge field $U(1)_X$ becomes massive via a St\"uckelberg mechanism as a consequence
of a Green-Schwarz (GS) anomaly cancellation~\cite{Green:1984sg,Dine:1987bq}, which is achieved through the coupling of
twisted Ramond-Ramond axions~\cite{Sagnotti:1992qw, Ibanez:1998qp}. The mass of the
anomalous~\footnote{Note that the $U(1)$ is not necessarily anomalous in four dimensions. A mass can be generated for a non-anomalous $U(1)$ by a six-dimensional (6d) GS term associated to a 6d anomaly cancellation in a sector of the theory.} $U(1)_X$ can be unambiguously
calculated by a direct one-loop string computation. Assuming the $U(1)_X$ arises from a brane wrapping $\delta_X$ dimensions among the $d$ large dimensions, it is given by
\begin{equation}
  m_X = \varkappa \ \sqrt{\frac{V_a M_s^{2}}{V_X M_s^{\delta}}} M_s =  \frac{\varkappa}{(2 \pi)^{\frac{\delta_x -2}{2}}} 
  \left( \frac{\sqrt{8}}{g_s} \frac{M_s}{M_{\rm Pl}} \right) ^{\frac{\delta_X- \delta_a}{d}} M_s\,, 
\end{equation}
where $\varkappa$ is the anomaly coefficient (which is in general an
ordinary loop suppressed factor), $V_a$ is the two-dimensional
internal volume corresponding to the propagation of the axion
field~\cite{Antoniadis:2002cs} and $ \delta_a$ is the number of large dimensions in $V_a$.  For $ \delta_a =0$, it  leads to:
\begin{equation}
  m_X =    \  \frac{\varkappa}{(2 \pi)^{\frac{\delta_x -2}{2}}} 
  \left( \frac{\sqrt{8}}{g_s} \frac{M_s}{M_{\rm Pl}} \right) ^{\delta_X/d} M_s\,  , 
\end{equation}
which gives $\sim  0.8 \varkappa$~keV and $\sim  38 \varkappa$~keV and
for $M_s \sim 10$~TeV and $M_s \sim 100$~TeV, respectively
($\delta_X=4$, $d=6$ and $g_s=0.2$). For this region of the parameter space, the gauge coupling
varies in the range $6 \times 10^{-10} \alt g_X \alt 6 \times
10^{-9}$. The case $\delta_a =2$ and $ \delta_X =d=4$ leads to:
\begin{equation}
  m_X =  \frac{\varkappa}{(2 \pi)} 
  \left( \frac{\sqrt{8}}{g_s} \frac{M_s}{M_{\rm Pl}} \right)
  ^{\frac{1}{2}} M_s \sim  172\  \varkappa  \left( \frac{0.2}{g_s}
  \right) ^{\frac{1}{2}}  \left(  \frac{M_s}{10\  {\rm TeV}} \right)
  ^{\frac{3}{2}} {\rm keV} \, .
\end{equation}
For a concrete example of such case, consider 2
D7-branes intersecting in two common directions; namely, $D7_1\!:1234$
and $D7_2\!:1256$, where $123456$ denote the internal six
directions. Take now 1234 large and 56 small (order the string scale)
compact dimensions. The gauge fields of $D7_1$ have a suppression of
their coupling by the 4-dimensional internal volume $V_X$ while the
states in the intersection of the two D7 branes see only the 12 large
dimensions and give 6 dimensional anomalies, cancelled by an axion
living in the same intersection, so $V_a$ is the volume of 12
only. 

 \subsection{$U(1)$ kinetic mixing}

We have seen that the tiny couplings are not trivial to obtain and lead often to too small dark photon masses.  This issue can be alleviated by resorting to the case where effective smaller couplings of $U(1)_X$ to SM states are obtained when the dark photons do not couple directly to the visible sector, but do it through kinetic mixing with ordinary photons.
 It can be generated by non-renormalisable operators, but it is natural to assume that it is generated by loops of states carrying charges $(q^{(i)}, q_X^{(i)})$ under the two $U(1)$'s and having masses $m_i$:
\begin{equation}
\epsilon_{\gamma X} =\frac{e g_X  }{16 \pi^2} \sum_i q^{(i)}
q_X^{(i)} \ln {\frac{ m_i^2}{{\mu^2}}} \equiv \frac{e g_X }{16 \pi^2}
C_{\rm Log} 
\end{equation}
where $\mu^2$ denotes the renormalization scale\footnote{In string theory, it is replaced by the string scale $M_s$.}, where we absorbed also the constant contribution. The effective coupling to SM is then: 
\begin{equation}
g_{X, {\rm eff}} = e \epsilon_{\gamma X}=   \frac {\alpha_{\rm em} g_X }{4 \pi}
C_{\rm Log} \sim 6 \times 10^{-4} g_X C_{\rm Log} \,.
\end{equation}
We can try to fit both desired values of $g_{X, {\rm eff}} $ and $m_X$. For a mass of the dark photon arising from a Higgs mechanism, we determine $g_x \sim m_X/M_s$, with $v_X \sim M_s$, this constrains:
\begin{eqnarray}
C_{\rm Log}  & \simeq & 1.7 \times 10^{3} \ g_{X, {\rm eff}} \ \frac{M_s}{m_X}
                        \nonumber \\
  & \simeq & 0.05 \ \left(\frac{g_{X, {\rm eff}}}{8\times 10^{-16}}
             \right) \
             \left(\frac{M_s}{100 \ {\rm TeV}}\right)
           \  \left(\frac{m_X}{2.8  \ {\rm keV}} \right)^{-1} \, .
\end{eqnarray}
A cancellation in the logarithm can be total, and the contribution appears at higher loops \cite{Gherghetta:2019coi}, or partial, for instance between particles with (order one) charges $(q^{(i)}, q_X^{(i)})$ and $(q^{(j)}, q_X^{(j)}= - q_X^{(i)})$ and masses $m_j=m_i + \Delta m_{ij}$. For $\Delta m_{ij} \ll m_i$, we have an approximation:
\begin{equation}
C_{\rm Log} \sim  \sum_{i,j} {\frac{ \Delta m_{ij}}{{m_i}}} \, .
\label{degenerescence}
\end{equation}

 \section{Towards explicit models}
 
We shall now discuss more explicitly the emergence of such extra abelian gauge groups in $D$-brane models.
The minimal embedding of the SM particle spectrum requires at least
three brane stacks~\cite{Antoniadis:2000ena} leading to three distinct
models of the type $U(3) \otimes U(2) \otimes U(1)$ that were
classified in~\cite{Antoniadis:2000ena,Antoniadis:2004dt}. Only one of
them, model {\it (C)} of~\cite{Antoniadis:2004dt}, has baryon number as a gauge symmetry that guarantees proton stability (in perturbation theory), and can be used in the framework of low mass scale string compactifications. In addition, because the charge associated to the $U(1)$ of $U(2)$ does not participate in the hypercharge combination, $U(2)$ can be replaced by the symplectic $S\!p(1)$ representation of Weinberg-Salam $S\!U(2)_L$, leading to a model with one extra $U(1)$ added to the hypercharge~\cite{Berenstein:2006pk}.
Note that the abelian factor associated to the $U(2)$ stack of
D-branes couples to the lepton doublet, and consequently this
anomalous $U(1)$ cannot be a good dark matter candidate, because the
left-handed neutrinos make it unstable. One can add to these three
stacks another D9-brane which will provide the  $U(1)_X$ which will
mix with the photon through loops of states living in the
intersections of the D9 and the  $U(3)$ and $U(1)$
stacks. The dark $U(1)_X$ is of course unstable as it decays
to three ordinary photons. However, the partial decay width is found to be~\cite{Redondo:2008ec}
\begin{equation}
    \Gamma_{X \to 3\gamma} \sim  10^{-28} \left(\frac{m_X}{2.8~{\rm keV}}
      \right)^9 \left(\frac{g_{X, {\rm eff}}}{5 \times 10^{-16}} \right)^2~{\rm Gyr}^{-1} \, ,
\end{equation}
and so for the range of small gauge
  coupling considered here, the life-time is big enough to allow it be
  a viable candidate for dark matter. 

Actually, the SM embedding in four D-brane stacks leads to many more models that
have been classified
in~\cite{Antoniadis:2002qm,Anastasopoulos:2006da}. The total gauge group of interest
here, 
\begin{eqnarray}
  G  & = & U(3)_C \otimes U(2)_L \otimes U(1)_1 \otimes U(1)_X  \\
  & = & S\!U(3)_C \otimes U(1)_C \otimes S\!U(2)_L  \otimes U(1)_L \otimes
        U(1)_1 \otimes U(1)_X \,, \nonumber 
\end{eqnarray}
contains four abelian factors. The non-abelian structure determines
the assignments of the SM particles. The quark doublet $Q$ corresponds to an open
string with one end on the color stack of D-branes and the other on the weak stack. The anti-quarks $u^c$ and $d^c$ must have one of their ends
attached to the color branes. The lepton doublet and possible Higgs
doublets must have one end on the weak set of branes.  Per contra, the
abelian structure is not fixed because the $U(1)_Y$ boson, which gauges the usual electroweak hypercharge symmetry,
could be a linear combination of all four abelian factors. However,
herein we restrict ourselves to models in which the bulk $U(1)_X$ does
not contribute to the hypercharge, in order to avoid an
unrealistically small gauge coupling. Of particular interest here are
models {\it (3)} and {\it (5)} of reference~\cite{Antoniadis:2002qm}. The general properties of their chiral
spectra are summarized in Table~\ref{tabla:1} and \ref{tabla:2}. One can check by inspection that for both models the hypercharge,
\begin{eqnarray}
  q_Y = -\frac{1}{3} q_C + \frac{1}{2} q_L + q_1 &~~~~{\rm for \ model}~\textit{(3)} \nonumber \\
  q_Y = \phantom{-}\frac{2}{3} q_C + \frac{1}{2} q_L + q_1 &~~~~{\rm for \
    model}~\textit{(5)} 
\end{eqnarray}
is anomaly free. In addition, the $U(1)_X$ is long-lived (because it only couples to the $e^c$ and to
either $u^c$ or $d^c$) and therefore a viable DM candidate.

\begin{table}
  \caption{Chiral fermion spectrum of the D-brane model {\it (3)}. \label{tabla:1}}
  \begin{tabular}{ccccccc}
    \hline
    \hline
    ~~Fields~~ & ~~Representation~~ &  ~~$q_C$~~ & ~~$q_L$~~ & ~~$q_1$~~ & ~~$q_X$~~ & ~~$q_Y$~~ \\
    \hline
    $Q$ & $\bm{(3,2)}$ & $\phantom{-}1$ & 1 & $\phantom{-}0$ & $\phantom{-}0$ & $\phantom{-}\frac{1}{6}$ \\
    $u^c$ & $\bm{(}\bar{\bm{3}},\bm{1)}$ & $-1$ & 0 & $-1$ & $\phantom{-}0$ & $-\frac{2}{3}$ \\ 
    $d^c$ & $\bm{(}\bar{\bm{3}},\bm{1)}$ & $-1$ & 0 & $\phantom{-}0$  & $-1$ & $\phantom{-}\frac{1}{3}$ \\
     $L$ & $\bm{(1,2)}$ & $\phantom{-}0$ & 1 &  $-1$  & $\phantom{-}0$   & $-\frac{1}{2}$\\
$e^c$ & $\bm{(1,1)}$ & $\phantom{-}0$     & 0  &  $\phantom{-}1$ & $\phantom{-}1$  & $\phantom{-}1$ \\
    \hline
    \hline
  \end{tabular}
  \end{table}

  \begin{table}
    \caption{Chiral fermion spectrum of the D-brane model {\it (5)}. \label{tabla:2}}
  \begin{tabular}{ccccccc}
    \hline
    \hline
    ~~Fields~~ & ~~Representation~~ &  ~~$q_C$~~ & ~~$q_L$~~ & ~~$q_1$~~ & ~~$q_X$~~ & ~~$q_Y$~~ \\
    \hline
    $Q$ & $\bm{(3,2)}$ & $\phantom{-}1$ & $-1$ & $\phantom{-}0$ & $0$ & $\phantom{-}\frac{1}{6}$ \\
    $u^c$ & $\bm{(}\bar{\bm{3}},\bm{1)}$ & $-1$ & $\phantom{-}0$ & $\phantom{-}0$ & $1$ & $-\frac{2}{3}$ \\ 
    $d^c$ & $\bm{(}\bar{\bm{3}},\bm{1)}$ & $-1$ & $\phantom{-}0$ & $\phantom{-}1$  & $0$ & $\phantom{-}\frac{1}{3}$ \\
     $L$ & $\bm{(1,2)}$ & $\phantom{-}0$ & $\phantom{-}1$ &  $-1$  & $0$   & $-\frac{1}{2}$\\
$e^c$ & $\bm{(1,1)}$ & $\phantom{-}0$     & $\phantom{-} 0$  &  $\phantom{-}1$ & $1$  & $\phantom{-}1$ \\
    \hline
    \hline
  \end{tabular}
  \end{table}

 \section{Summary}

We have investigated the possibility of identification of the light dark photon  with one of
the ubiquitous $U(1)$ gauge bosons of D-brane string theory constructions. We have first investigated how small 
the gauge coupling can be made. We found that open strings allow values as weak as $g_X \sim \mathcal{O}(10^{-14})$ for a string scale 
of order $M_s \sim \mathcal{O}(10)$ TeV. For the case of six and four large extra dimensions, 
the Kaluza-Klein excitations appear above the GeV and MeV scales, respectively. They are
very weakly coupled and decay quickly, but one could hope to observe their cumulative effect at TeV scales energies in future collider searches. 
The case  of small instantons leads to similar conclusions 
with an interesting additional possibility: no large extra dimension below the TeV but a tiny string coupling. This possibility is also realized in 
Little String Theory but with a stronger gauge coupling $g_X \sim \mathcal{O}(10^{-7})$. We have then looked at possible realization of models with dark photon 
masses in the range of  keV. We have found two possibilities. A stringy St\"uckelberg mechanism generates masses in the desirable range 
in models with four large extra dimensions. If the dark photon mass is generated instead by a (low energy field theoretical) Higgs mechanism, 
then a kinetic mixing between the visible and dark $U(1)$  allows to get simultaneously the desired mass and coupling strength.

Here, we have left aside a couple of issues whose investigations are beyond the scope of this work. String models 
typically exhibit a large number of moduli fields that need to be fixed. In particular, 
the size of extra dimensions and couplings discussed here are vacuum expectation of such fields and all the hierarchies should be dynamically 
generated (see \cite{Cicoli:2011yh} for example). Also, obtaining the correct relic density for such weakly coupled particles 
to form the main component of dark matter is challenging. The dark photons can not be thermally produced and one needs to resort to different mechanisms. 
A promising possibility is the proposal of \cite{Graham:2015rva}, the desired abundance of dark photons can be generated by the quantum fluctuations during inflation 
 for an appropriate set of parameters: the scale of inflation, reheating temperature and coupling to the curvature scalar.

%
%
%
%

Finally, we have discussed possible D-brane model can accommodate the
excess of events with $3\sigma$ significance over background recently
observed at XENON1T. The model is fully predictive, and can be
confronted with future data from dark matter direct-detection
experiments, LHC Run 3 searches,  and astrophysical observations. \\

The work of L.A.A. is supported  by the U.S. National Science Foundation (NSF Grant PHY-1620661) and
  the National Aeronautics and Space Administration (NASA Grant
  80NSSC18K0464). The research of I.A. is funded in part by the
  ``Institute Lagrange de Paris'', and in part by a CNRS PICS grant. The work of K.B. is
   supported by the Agence Nationale de Recherche under grant ANR-15-CE31-0002 ``HiggsAutomator''. The work of D.L. is
  supported by the Origins Excellence Cluster.  Any opinions, findings, and
  conclusions or recommendations expressed in this material are those
  of the authors and do not necessarily reflect the views of the NSF
  or NASA.

\end{document}